\documentstyle[preprint,prd,aps]{revtex}
\newcommand{\be}{\begin{equation}}
\newcommand{\ee}{\end{equation}}
\newcommand{\bea}{\begin{eqnarray}}
\newcommand{\eea}{\end{eqnarray}}
\newcommand{\sptwo}{1.4}
\newcommand{\doublespace}{\edef\baselinestretch{\sptwo}\Large\normalsize}
\newcommand{\newsection}[1]{\setcounter{equation}{0}}

\newcounter{newapp}
\begin{document}
\begin{titlepage}
\vspace*{1.0in}
\begin{center}
{\bf Gauging  internal fermionic symmetries and spin 3/2 fields}
\end{center}
\begin{center} 
S.T. Love\footnote{e-mail address: love@physics.purdue.edu} \\
{\it Department of Physics\\ 
Purdue University\\
West Lafayette, IN 47907-2306}
~\\
\end{center}
\vspace{1in}
\begin{center}
{\bf Abstract}\\
Field theoretic models possessing a global internal fermionic shift symmetry are considered. When such a symmetry is realized locally, spin 3/2 fields appear naturally  as gauge fields. Implementation of the gauging procedure requires not only the usual replacement of ordinary derivatives by covariant derivatives containing the spin 3/2 fields, but also the inclusion of additional monomials. 
The Higgs mechanism and the high energy Nambu-Goldstone fermion equivalence theorem are 
explicitly demonstrated. 
\end{center}

\end{titlepage}

\doublespace

\underline\\

Spin 1 fields arise most naturally as the vector connections when global internal symmetries are realized as local symmmetries. This gauging of global symmetries and the resultant Higgs mechanism is at the heart of the much of particle physics including the Standard Model. In this note, we address the possibility of the origin of spin 3/2 particles as the result of gauging a global internal fermionic symmetry. The most familiar fermionic symmetry is supersymmetry (SUSY), a spacetime symmetry related through the SUSY algebra to spacetime translations. Thus gauging supersymmetry automatically requires a gauging of the full Poincar\'e group resulting in a supergravity theory which, in addition to the spin 3/2 fields, necessarily contains the spin 2 graviton field. Here we examine models with a global internal fermionic symmetry whose gauging can be implemented only using the spin 3/2 fields without the need for any other higher spin fields. 
 
We begin by introducing a model which contains a 2-component Weyl fermion, $\lambda^\alpha$ and its complex conjugate $\bar{\lambda}_{\dot{\alpha}}$ and demand the model be invariant under the global shift transformation:
\bea
\delta(\xi, \bar{\xi})\lambda &=& f^{3/2}\xi \nonumber\\
\delta(\xi, \bar{\xi})\bar{\lambda} &=& f^{3/2}\bar{\xi}
\eea
where $\xi, \bar{\xi}$ are spacetime independent Grassmann parameters and $f$ is a mass scale. The resulting invariant action, retaining terms through mass dimension 4 is just the kinetic term of the free Dirac action:
\be
I_0=\int d^4x {\cal L}_0 ~~;~~
{\cal L}_0 =-\frac{i}{2}\lambda \sigma^\mu\stackrel{\leftrightarrow}\partial_\mu \bar{\lambda}
\label{FDA}
\ee
Note that the required symmetry forbids a $\lambda\lambda +\bar{\lambda}\bar{\lambda}$ fermion mass term. 

The globally symmetric model can be readily extended to include higher dimensional (derivatively coupled) fermionic self interations. In fact, any term of the form
\be
{\cal L}_1 = \lambda \sigma^\mu \partial_\mu 
\bar{F}_1(\partial \lambda, \partial \bar\lambda) + h.c.
\ee
where $\bar{F}_1^{\dot\alpha}(\partial \lambda, \partial \bar\lambda)$ is a function of $\partial\lambda$ and $\partial \bar\lambda$ produces an invariant action. An example of such an operator is provided by $\bar{F}_1^{\dot\alpha} =\frac{a_1}{f^2}\partial^2 \bar\lambda^{\dot\alpha} $, where $a_1$ is a constant.

The model can be further extended to include couplings to other fields (generically referred to as $\phi_i$). Under the global shift of the fermion field, all these other fields are taken to be invariant: $\delta (\xi,\bar\xi)\phi_i =0$. Thus a globally invariant Lagrangian containing no fermion fields experiencing the nontrivial shift is simply
\be
{\cal L}_2=F_2(\phi_i,\partial\phi_i)
\ee
This could include the full Standard Model Lagrangian. \\
One can also couple the fields $\phi_i$ to the $\lambda, \bar{\lambda}$ with an invariant Lagrangian of the form
\be
{\cal L}_3= F_3(\phi_i,\partial\phi_i,\partial\lambda,\partial\bar{\lambda})
\ee

Once again focusing on the free massless fermion field theory exhibiting the global shift symmetry, we now promote the Grassmann symmetry parameters to become local, (local variations and locally invariant actions and their associated Lagrangians will be denoted using a hat) $\xi \rightarrow \xi(x)~~;~~\bar{\xi}\rightarrow \bar{\xi}(x)$, so that 
\bea
\hat{\delta}(\xi, \bar{\xi})\lambda &=& f^{3/2}\xi(x) \nonumber\\
\hat{\delta}(\xi, \bar{\xi})\bar{\lambda} &=& f^{3/2}\bar{\xi}(x)
\eea
We seek an action generalizing that of Eq. (\ref{FDA}) which is invariant under this local transformation. The standard procedure is to simply replace the derivatives by covariant derivatives
\bea
\partial_\mu\lambda &\rightarrow & D_\mu\lambda =\partial_\mu\lambda+if \psi_\mu \nonumber \\
\partial_\mu\bar{\lambda} &\rightarrow & D_\mu\bar{\lambda} =\partial_\mu\bar{\lambda}-if \bar{\psi}_\mu 
\eea
where $\psi_\mu^\alpha $ and its conjugate $ \bar{\psi}_{\dot{\alpha}}^\mu $ are spin 3/2 fields carrying both vector and spinor indices. Making such a substitution, the Lagrangian of Eq. (\ref{FDA}) takes the form
\be
{\cal L}_0=-\frac{i}{2}\lambda \sigma^\mu\stackrel{\leftrightarrow}D_\mu\bar{\lambda}=-\frac{i}{2}\lambda \sigma^\mu\stackrel{\leftrightarrow}\partial_\mu \bar{\lambda} -\frac{f}{2}(\lambda \sigma^\mu \bar{\psi}_\mu +\psi_\mu\sigma^\mu \bar{\lambda})
\ee
The local $\xi$ variation of this Lagrangian (the $\bar{\xi}$ variation can be obtained by complex conjugation) is then 
\be
\hat{\delta}(\xi){\cal L}_0 = -\frac{i}{2}[f^{3/2}\xi\sigma^\mu \partial_\mu \bar{\lambda}-f^{3/2}\partial_\mu \xi \sigma^\mu -f^{5/2}\xi \sigma^\mu \bar{\psi}_\mu -if \lambda\hat{\delta}(\xi)(\sigma^\mu\bar{\psi}_\mu) -if(\hat{\delta}(\xi)\psi_\mu) \sigma^\mu \bar{\lambda}]
\ee
As usual, we can take 
\be
\hat{\delta}(\xi)\psi_\mu = if^{1/2}\partial_\mu\xi
\ee
to cancel the part of the variation depending on $\partial_\mu\xi$. However, there is nothing to cancel the terms dependent on $\xi$ (not derivative of $\xi$). The source of the dilema is the fact that the Dirac Lagrangian is not a function of only derivatives of the fermion fields. This case can be compared to the massless scalar kinetic term which is a function of only derivatives of the scalar fields. There simply replacing ordinary derivatives by covariant derivatives which include the spin 1 vector field is sufficient to render that model locally invariant. 

Thus to construct a locally invariant model for the fermion case, we must do more than just replace derivatives by covariant derivatives. We need to include additional terms in the action as well as have something which transforms into $\xi$ and not its derivative. We already have that $\psi_\mu$ transforms into a derivative of $\xi$. The only other object remaining is $\bar{\psi}_\mu$. Recall that the spin 1/2 field $\lambda$ transforms as $\xi$. Thus we demand that spin 3/2 field $\bar{\psi}_\mu$ transform as
\be
\hat{\delta}(\xi)\bar{\psi}^\mu =-\frac{1}{4}f^{3/2}\bar{\sigma}^\mu \xi 
\ee
so that $ \hat{\delta}(\xi)(\sigma^\mu\bar{\psi}_\mu ) = f^{3/2}\xi $.
 
To proceed, we next systematically list the operators and their associated variations starting with the Dirac covariant kinetic term for $\lambda$ and $\bar{\lambda}$ and then cancelling the unwanted variations until we left with an invariant action under the transformations
\bea
\hat{\delta}(\xi)\lambda &=&f^{3/2} \xi \nonumber \\
\hat{\delta}(\xi)\bar{\lambda} &=& 0 \nonumber \\
\hat{\delta}(\xi)\psi_\mu &=& f^{1/2} \partial_\mu \xi \nonumber \\
\hat{\delta}(\xi)\bar{\psi}^\mu &=& -\frac{1}{4}f^{3/2} \bar\sigma^\mu \xi 
\label{TLL}
\eea
and the complex conjugate (i.e. $\bar{\xi}$) variations. 

The needed operators and their local transformations are given as:
\bea
{\cal O}_1 &=& -\frac{i}{2} \lambda \sigma^\mu \stackrel{\leftrightarrow}{\cal D}_\mu \bar{\lambda}~~;~~\hat{\delta}(\xi){\cal O}_1= -\frac{1}{2}f^{3/2} (i\xi\sigma^\mu {\cal D}_\mu \bar{\lambda}+f\xi\lambda )\nonumber \\
{\cal O}_2 &=&f(\lambda \lambda +\bar{\lambda}\bar{\lambda})~~;~~\hat{\delta}(\xi){\cal O}_2= 2f_S^{5/2} (\xi\lambda )\nonumber \\
{\cal O}_3 &=& \frac{1}{f}({\cal D}_\mu\lambda {\cal D}^\mu \lambda +{\cal D}_\mu\bar{\lambda}{\cal D}^\mu\bar{\lambda}  -2{\cal D}_\mu \lambda \sigma^{\mu\nu}{\cal D}_\nu \lambda -2{\cal D}_\mu\bar{\lambda}\bar{\sigma}^{\mu\nu}{\cal D}_\nu\bar{\lambda} ) ~~;~~\hat{\delta}(\xi){\cal O}_3= \frac{1}{2}f^{3/2}(i\xi \sigma^\mu{\cal D}_\mu \bar{\lambda})\nonumber \\
&&
\eea
Next form the Lagrangian
\be
\hat{\cal L}_0 = \sum_{i=0}^3 c_i {\cal O}_i
\ee
and demand that its action is invariant under the variations. This leads to  relationships among the various coeficients $c_i$. Canonically normalizing the kinetic term so that $c_1 =1$, then fixes $c_2 =\frac{1}{4}~~;~~c_3 =-\frac{1}{4}$.

Thus the action, $\hat{I}_0$, formed from the Lagrangian \\
\be 
\hat{\cal L}_0 = -\frac{i}{2}\lambda \sigma^\mu \stackrel{\leftrightarrow}{\cal D}_\mu \bar{\lambda} +\frac{f}{4}(\lambda \lambda +\bar{\lambda}\bar{\lambda})-\frac{1}{4f}
({\cal D}_\mu \lambda {\cal D}^\mu \lambda + {\cal D}_\mu \bar{\lambda} {\cal D}^\mu \bar{\lambda} -2 {\cal D}_\mu \lambda \sigma^{\mu\nu}{\cal D}_\nu \lambda -2 {\cal D}_\mu \bar{\lambda} \bar{\sigma}^{\mu\nu}{\cal D}_\nu \bar{\lambda})
\label{LSA}
\ee
is invariant under the local variations: 
$\hat{\delta}(\xi, \bar{\xi})\hat{I} =0$. Included in the product of covariant derivatives in $\hat{\cal L}$ is a mass term for the spin 3/2 field of the form  $\frac{m_{3/2}}{2}(\psi^\mu\psi_\mu +\bar{\psi}^\mu\bar{\psi}_\mu )$ with $m_{3/2}=\frac{f}{2}$. It follows that a Higgs mechanism has taken place and the Nambu-Goldstone spin 1/2 fermion fields ($\lambda$ and $\bar{\lambda}$) become the spin 1/2 components of the massive spin 3/2 fields. This is also consistent with the transformation laws $\hat{\delta}(\xi)\psi_\mu = if^{1/2} \partial_\mu \xi ~~;~~\hat{\delta}(\sigma^\mu\bar{\psi}_\mu )= f^{3/2} \xi $. In contrast to the scalar case where the vector mass term emerges directly from the boson kinetic term when derivatives are replaced with covariant derivatives, the appearance of spin 3/2 field mass term is somewhat more circuitous. It arises from an independent monomial needed to be added to the action in order to inforce the local shift invariance. Note that this Lagrangian is invariant under the local fermionic shift transformation but violates the global chirality of the globally shift invariant theory: $\delta (\omega)\lambda=i\omega\lambda ~~;~~\delta (\omega)\bar{\lambda}=-i\omega\bar{\lambda}$. In particular, the Nambu-Goldstone spin 1/2 fermion fields ($\lambda$ and $\bar{\lambda}$) acquire a mass term of the form $\frac{m_{3/2}}{2}(\lambda \lambda +\bar{\lambda}\bar{\lambda})$ where the coefficient is identical to the mass of the spin 3/2 field; $m_{3/2}=f/2$.

To complete the program, we need to construct the locally invariant kinetic term for the spin 3/2 fields. It is readily established that the Lagrangian 
\be
\hat{\cal L}_{RS} = -\frac{i}{2}\psi_\mu (\sigma^\nu \stackrel{\leftrightarrow}\partial_\nu \bar{\psi}^\mu -\sigma^\mu \stackrel{\leftrightarrow}\partial_\nu \bar{\psi}^\nu +i\epsilon^{\mu\nu\lambda\rho}\sigma_\nu \stackrel{\leftrightarrow}\partial_\lambda\bar{\psi}_\rho )
\ee
transforms as a total divergence under the local variations given by Eq. (\ref{TLL}). 
Consequently, the (Rarita-Schwinger) action\cite{RS} $\hat{I}_{RS}=\int d^4x \hat{\cal L}_{RS}$ formed from this Lagrangian is locally invariant. 

The total locally invariant action, $\hat{I}$, is obtained by adding $\hat{I}_{RS}$ to the action formed from Eq. (\ref{LSA}): $\hat{I}=\hat{I}_0+\hat{I}_{RS}$. The dynamical degrees of freedom can be immediately gleaned by going to unitary gauge which is tantamount to setting $\lambda =0= \bar{\lambda}$ . So doing, one finds 
\bea
\hat{\cal L}_{\rm unitary ~gauge} &=& -\frac{i}{2}\psi_\mu (\sigma^\nu \stackrel{\leftrightarrow}\partial_\nu \bar{\psi}^\mu -\sigma^\mu \stackrel{\leftrightarrow}\partial_\nu \bar{\psi}^\nu +i\epsilon^{\mu\nu\lambda\rho}\sigma_\nu \stackrel{\leftrightarrow}\partial_\lambda\bar{\psi}_\rho ) \nonumber \\
&&+\frac{f}{4}[\bar{\psi}_\mu \bar{\psi}^\mu +\psi_\mu \psi^\mu -2\bar{\psi}_\mu \bar{\sigma}^{\mu\nu}\bar{\psi}_\nu -2\psi_\mu\sigma^{\mu\nu}\psi_\nu ]
\eea
which is the Lagrangian of a massive spin 3/2 field with mass term $m_{3/2}=\frac{f}{2}$. This is the Lagrangian relevant for the infrared dynamics. 

On the other hand, for energies much greater than the mass of the spin 3/2 particles, the model obeys an equivalence theorem which equates the scattering amplitudes of the (longitudnal polarized) spin 1/2 components of the spin 3/2 particles to the scattering amplitudes of the Nambu-Goldstone fermions. This is the analogous relation to the familiar high energy equivalence theorem between the scattering amplitudes of longitudnally polarized vector gauge bosons and the associated Nambu-Goldstone bosons\cite{NGBET}. To derive the Nambu-Goldstone fermion equivalence theorem, consider the generating functional of connected Green functions for the spin 1/2 components of the spin 3/2 gauge fermions given by
\be
W[\eta^{(\pm \frac{1}{2})}]=-i \ell n\int [d\psi_\mu^\alpha][d\bar\psi_\mu^{\dot\alpha}][d\lambda^\alpha][d\bar\lambda^{\dot\alpha}]e^{i\hat{I}}e^{i\int d^4x [\sum_{\sigma = \pm\frac{1}{2}} \eta^{(-\sigma)}\psi^{(\sigma)}+ h.c.]}\delta[F(\psi,\lambda)]\delta[\bar{F}(\bar\psi,\bar\lambda)]
\ee
where $\hat{I}$ is the locally invariant action derived above. The gauge fixing (spinor) functions\cite{BGO} are taken as 
\bea
F^\alpha(\psi,\lambda)&=&(\sigma^\nu\partial_\nu \bar\sigma^\mu\psi_\mu)^\alpha +\sqrt{\frac{3}{2}}m_{3/2}\lambda^\alpha \nonumber \\
\bar{F}_{\dot\alpha}(\bar\psi,\bar\lambda)&=&(\bar\sigma^\nu\partial_\nu\sigma^\mu\bar\psi_\nu)_{\dot\alpha}+ \sqrt{\frac{3}{2}}m_{3/2}\bar\lambda_{\dot\alpha}
\eea
Note that with this gauge choice, no ghost fields are required.
The $\eta^{(\pm\frac{1}{2})}$ are the (Grassmann) sources for the on mass shell spin 1/2 (longitudnal) components of the spin 3/2 fields which are given in momentum space using the vector-spinor wavefunctions\cite{AB} as 
\be
\tilde\psi^{(\pm\frac{1}{2})}(k)=\frac{i}{\sqrt{3}}[\epsilon_\mu^{(\pm 1)}(k) u^{(\mp \frac{1}{2})\alpha }(k)+\sqrt{2}\epsilon^{(0)}_\mu(k)u^{(\pm\frac{1}{2})^\alpha }(k)]\tilde\psi^\mu_\alpha (k)
\ee
with $k^\mu =(\sqrt{\vec{k}^2+m_{3/2}^2},\vec{k})$  being the 4-momentum carried by the spin 3/2 fermion. Here $\epsilon_\mu^{(\pm1)}(k)$ and $\epsilon_\mu^{(0)}(k)$ are the spin-1 transverse  and longitudnal spin-1 polarization vectors respectively,  $u^{(\pm\frac{1}{2})}(k)$ are the 2-component spinors corresponding to the spin polarizations $\pm\frac{1}{2}$ and $\tilde\psi^\mu_\alpha(k)$ is the Fourier transform of the spin 3/2 field $\psi_\mu^\alpha(x)$. At high energies, $|\vec{k}|>>m_{3/2}$, the longitudnal spin-1 polarization vector, $\epsilon_\mu^{(0)}(k)=\frac{1}{m_{3/2}}(|\vec{k}|, \sqrt{\vec{k}^2+m_{3/2}^2}\hat{k})$,  grows as $\epsilon^{(0)}_\mu (k) = \frac{k_\mu}{m_{3/2}}+{\cal O}(\frac{m_{3/2}}{|\vec{k}|})$ and hence gives the dominant contribution. It follows that in this limit
\be
\tilde\psi^{(\pm\frac{1}{2})}(k)=-i(\frac{2}{3})^{1/2}\frac{k^\mu}{m_{3/2}}u^{(\pm\frac{1}{2})^\alpha }(k)\tilde\psi^\mu_\alpha (k)+ {\cal O}(\frac{m_{3/2}}{|\vec{k}|})~~;~~|\vec{k}|>>m_{3/2}
\label{HEL}
\ee

Using the on mass shell conditions for the scattering amplitudes, the gauge fixing term
\be
(\sigma^\nu\partial_\nu\bar\sigma^\mu\psi_\mu)^\alpha =-\sqrt{\frac{3}{2}}m_{3/2}\lambda^\alpha
\ee
reduces upon application of the field equations to the constraints
\be
(\sigma^\mu \psi_\mu)_\alpha = -i \sqrt{\frac{3}{2}}\lambda_\alpha
\ee
and 
\be
\partial_\mu \psi^\mu_\alpha =\sqrt{\frac{3}{2}}m_{3/2}\lambda_\alpha
\label{EGC}
\ee
In obtaining these results, it was crucial that the mass, $m_{3/2}$, for the spin 3/2 fields, $\psi_\mu^\alpha, \bar\psi_\mu^{\dot\alpha}$ and spin 1/2 fields, $\lambda^\alpha
, \bar\lambda^{\dot\alpha}$, were identical. Using Eq.(\ref{EGC}) in  Eq. (\ref{HEL}) then gives 
\be
\tilde\psi^{(\pm\frac{1}{2})}(k) = u^{(\pm \frac{1}{2}) \alpha }(k)\tilde\lambda_\alpha (k) + {\cal O}(\frac{|\vec{k}|}{m_{3/2}})\equiv \tilde\lambda^{(\pm \frac{1}{2})}(k)
~~;~~|\vec{k}|>>m_{3/2}
\ee

Susbtituting this result into the generating functional for connected on-shell Green functions for the spin 1/2 components of the spin 3/2 gauge fermions allows it to be written at high energies, ($|\vec{k}|>>m_{3/2}$), as
\be
W[\eta^{(\pm \frac{1}{2})}]=-i \ell n \int [d\psi_\mu^\alpha][d\bar\psi_\mu^{\dot\alpha}][d\lambda^\alpha][d\bar\lambda^{\dot\alpha}]
e^{i\hat{I}}e^{i\int d^4k [\sum_{\sigma=\pm\frac{1}{2}}\tilde\eta^{(-\sigma)}(-k)\tilde\lambda^{(\sigma)}(k)+ h.c.]}\delta[F(\psi,\lambda)]\delta[\bar{F}(\bar\psi,\bar\lambda)]
\ee
which establishes the equivalence theorem. That is, at high energies, the (longitudnal projected) spin 1/2 helicity scattering amplitudes of the spin 3/2 particle can be computed using the Nambu-Goldstone fermion scattering amplitudes.

Finally, let us consider the locally invariant coupling of the spin 3/2 field to the locally invariant matter. To do so, note that the combinations
$\psi^\mu -\frac{1}{f}\partial^\mu\lambda+\frac{1}{4}\sigma^\mu\bar{\lambda}$ and $\bar{\psi}^\mu-\frac{1}{f}\partial^\mu\bar{\lambda}+\frac{1}{4}\bar{\sigma}^\mu\lambda$ are locally shift invariant. Thus any Lorentz invariant function of these combinations and the $\phi_i$ of the form
\be
\hat{\cal L}_{int}= \hat{\cal L}_{int}(\psi^\mu -\frac{1}{f}\partial^\mu\lambda+\frac{1}{4}\sigma^\mu\bar{\lambda},\bar{\psi}^\mu-\frac{1}{f}\partial^\mu\bar{\lambda}+\frac{1}{4}\bar{\sigma}^\mu\lambda,\phi_i, \partial_\mu\phi_i)
\ee
will be locally shift invariant. 

In this note, we have seen how spin 3/2 fields can arise quite naturally when a global internal fermionic shift symmetry is gauged. The gauging procedure, however, requires the inclusion of further monomials in addition to the replacement of ordinary derivatives by covariant derivatives containing the spin 3/2 field. Thus the procedure a bit more involved than that needed for global bosonic internal symmetries. Note that the analogous global shift symmetry in the free massless scalar model when gauged produces the $U(1)$ nonlinear sigma model. Further note that as opposed to the gauging of supersymmetry\cite{WB}, which is a global spacetime fermionic symmetry, there is no need to include any other higher dimensional fields beyond the spin 3/2 gauge field. 

In closing, let us point out that the models possessing the global internal fermionic shift symmetry which we have considered in this paper do not violate the Coleman-Mandula theorem\cite{CM} or its extension by Haag, Lopuszanski and M. Sohnius\cite{HLS} which states that of all graded Lie algebras, only the supersymmetry algebras (and their extensions to include scalar central charges) generate symmetries of the S-matrix consistent with relativistic quantum field theory. The proof of that theorem contains the condition that there is an energy gap between the vacuum and the one particle state. This condition is not satisfied by the model possessing the fermionic shift symmetry. \\
\\

\noindent
During the course of this study, I have enjoyed several useful conversations with T.E. Clark. This work was supported in part by the U.S. Department 
of Energy under grant DE-FG02-91ER40681 (Task B).


\begin{references}

\bibitem{RS}W. Rarita and J. Schwinger, {\it Phys. Rev.} {\bf 60}, 61 (1941).

\bibitem{NGBET}J.M. Cornwall, D.N. Levin and G. Tiktopoulos, {\it Phys. Rev D} {\bf 10}, 1145 (1974); C.E. Vayonakis, {\it Lett. Nuovo Cimento} {\bf 17}, 383 (1976); B.W. Lee, C. Quigg and H.B. Thacker, {\it Phys. Rev. D} {\bf 16}, 1519 (1977). For an analogous equivalence theorem equating the high energy scattering amplitudes of the spin 1/2 projections of the spin 3/2 gravitino to the scattering amplitudes of the Goldstino of spontaneously broken supersymmetry, see R. Casalbuoni, S. De Curtis, D. Dominici, F. Feruglioa and R. Gatto, {\it Phys. Lett.} {\bf B215}, 313 (1988); {\it Phys. Rev. D} {\bf 39:}, 2281 (1989).

\bibitem{BGO}L. Baulieu, A. Georges and S. Ouvry, {\it Nucl. Phys.} {\bf B273}, 366 (1986).

%
\bibitem{AB} P.R. Auvil and J.J. Brehm, {\it Phys. Rev.} {\bf 145}, 112 (1966).

\bibitem{WB} See, for example, J. Wess and J. Bagger, {\it Supersymmetry and 
Supergravity},  (Princeton University Press, Princeton 1992); H.P. Nilles, {\it Phys. Rept.} {\bf 110}, 1 (1984).

\bibitem{CM}S. Coleman and J. Mandula, {\it Phys. Rev.} {\bf 159}, 1251 (1967).

\bibitem{HLS}R. Haag, J. Lopuszanski and M. Sohnius, {\it Nucl. Phys. B}, 257 (1975).

\end{references}
\end{document}